\documentclass[conference]{IEEEtran}
\IEEEoverridecommandlockouts
\usepackage{cite}
\usepackage{amsmath,amssymb,amsfonts}
\usepackage{algorithmic}
\usepackage{graphicx}
\usepackage{textcomp}
\usepackage{xcolor}
\usepackage{fancyhdr}

\usepackage{caption2}
\usepackage{booktabs}
\usepackage{mathrsfs}
\usepackage{epsfig}
\def\BibTeX{{\rm B\kern-.05em{\sc i\kern-.025em b}\kern-.08em
    T\kern-.1667em\lower.7ex\hbox{E}\kern-.125emX}}
\begin{document}

\title{JALAD: Joint Accuracy- and Latency-Aware Deep Structure Decoupling for Edge-Cloud Execution
\thanks{This work is supported in part by NSFC under Grant No.~61872215 and 61531006.}
}

\author{
	\IEEEauthorblockN{Hongshan Li\IEEEauthorrefmark{1}, Chenghao Hu\IEEEauthorrefmark{2}, Jingyan Jiang\IEEEauthorrefmark{3}, 
		Zhi Wang\IEEEauthorrefmark{4}, Yonggang Wen\IEEEauthorrefmark{5}, Wenwu Zhu\IEEEauthorrefmark{6}}
	\IEEEauthorblockA{\IEEEauthorrefmark{1}\IEEEauthorrefmark{6}Tsinghua-Berkeley Shenzhen Institute, Tsinghua University}
	\IEEEauthorblockA{\IEEEauthorrefmark{2}\IEEEauthorrefmark{4}Graduate School at Shenzhen, Tsinghua University}
	\IEEEauthorblockA{\IEEEauthorrefmark{3}Department of Computer Science, Jilin University}
	\IEEEauthorblockA{\IEEEauthorrefmark{5}School of Computer Science and Engineering, Nanyang Technological University}
	\IEEEauthorblockA{\IEEEauthorrefmark{6}Department of Computer Science and Technology, Tsinghua University
	\\\{lhs17, huch16\}@mails.tsinghua.edu.cn, jiangjy14@mails.jlu.edu.cn
	\\wangzhi@sz.tsinghua.edu.cn, ygwen@ntu.edu.sg, wwzhu@tsinghua.edu.cn}
}

\maketitle

\begin{abstract}
	
	Recent years have witnessed a rapid growth of deep-network based services and applications. A practical and critical problem thus has emerged: how to effectively deploy the deep neural network models such that they can be executed efficiently. Conventional cloud-based approaches usually run the deep models in data center servers, causing large latency because a significant amount of data has to be transferred from the edge of network to the data center. In this paper, we propose \textbf{JALAD}, a joint accuracy- and latency-aware execution framework, which decouples a deep neural network so that a part of it will run at edge devices and the other part inside the conventional cloud, while only a minimum amount of data has to be transferred between them. Though the idea seems straightforward, we are facing challenges including i) how to find the best partition of a deep structure; ii) how to deploy the component at an edge device that only has limited computation power; and iii) how to minimize the overall execution latency. Our answers to these questions are a set of strategies in JALAD, including 1) A normalization based in-layer data compression strategy by jointly considering \emph{compression rate} and \emph{model accuracy}; 2) A latency-aware deep decoupling strategy to minimize the overall execution latency; and 3) An edge-cloud structure adaptation strategy that dynamically changes the decoupling for different network conditions. Experiments demonstrate that our solution can significantly reduce the execution latency: it speeds up  the overall inference execution with a guaranteed model accuracy loss.
	
\end{abstract}


\section{Introduction}
\label{sec:intro}

The recent couple of years have witnessed a rapid growth of deep learning driven applications. A variety of deep network structures have been developed for image recognition (e.g., Magvii Face++), nature language process (e.g., Apple Siri), intelligent personal assistance services (e.g., Google Now), etc. As such deep neural network based machine learning models are driving the online intelligent services, a critical problem, how to efficiently deploy such models to support their users' requests, has emerged.

Conventional deep neural network model deployment usually relies on a cloud-based paradigm, to execute all the deep structures inside the servers, usually located in a centralized data center \cite{google-tpu,apple-siri}. The limitation of such solution is that a user usually has to upload a large amount of data to the cloud and retrieve the results. For example, in a facial recognition service, a raw 24-bit $1024\times768$ image with size around $2.4$MB (resp. $\sim1$MB in PNG compression), needs $2.4$ seconds (resp. $1$ second) to be uploaded to the server even if the uplink bandwidth is $1$MBps. The situation is worse given that the average uplink capacity in today's Internet can be smaller than that, especially for mobile networks \cite{index520862global}. How to reduce the \emph{execution latency} has challenged conventional deep model deployment solutions.

An intuitive idea to reduce the execution latency is to move the models to the edge devices of the network (e.g., base stations nearby) \cite{roman2016mobile}, so that uploading the data to the edge devices is much faster. But the limited computation and storage capacities of edge devices are usually not able to fully deploy and run large deep network models (e.g., containing up to millions of parameters). 

Another solution is to run a mini-version of the original deep network models on edge devices, usually a compressed version of the original deep network model exported by pruning \cite{structured_pruning, pruning_han, prun_quanti} or quantization \cite{quantization, quantize_3bit, quantize}. But the accuracy loss of these mini-versions are fixed when it is retrained, users can hardly customize and adaptively update the trade-off of accuracy and running latency for various applications with different accuracy demand upon a retrained model.

To solve this problem, partitioning a pre-trained deep network structure, i.e., its topology and corresponding parameters, has been proposed. By strategical partitioning, one part can fit into the computation power of the edge devices \cite{Neurosurgeon}. However, due to the ``data amplification'' effect, i.e., the size of data can increase when passing through the layers in a deep network, usually caused by the deep structure including \emph{convolutional} layers, such partition actually fails to reduce the size of data to be uploaded, thus not able to speed up much.

\begin{figure*}[t]
	\centering
	\includegraphics[width=0.9\linewidth]{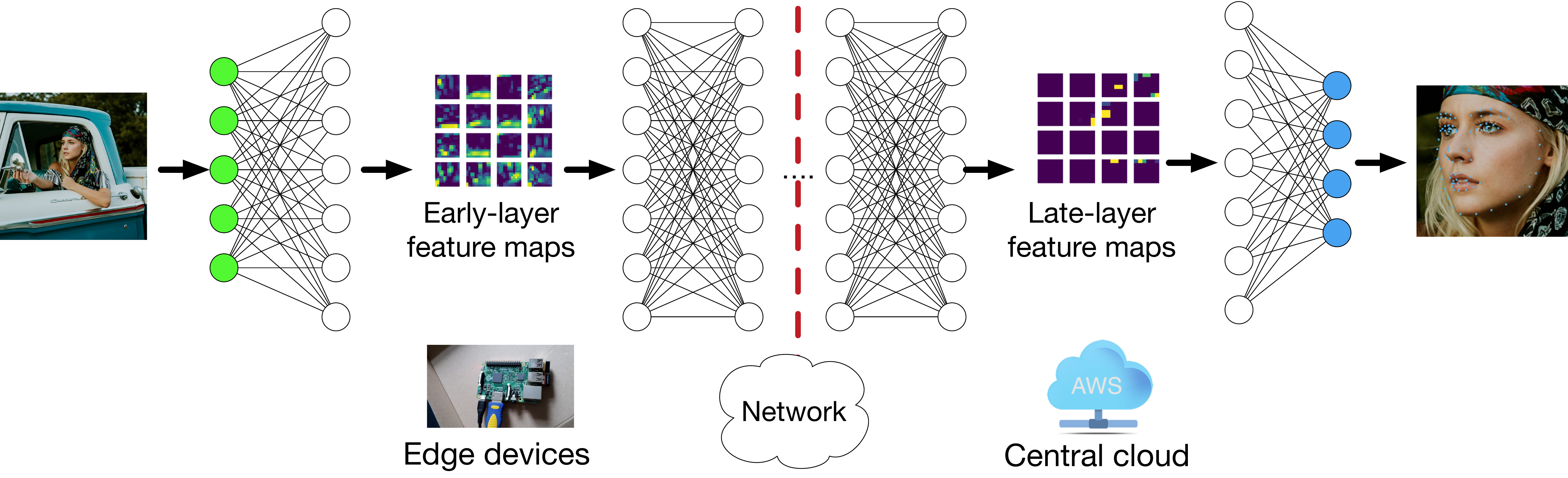}
	\caption{Framework of our joint accuracy- and latency-aware deep network decoupling, we find the optimal partition to run a deep neural network structure across edge device and central cloud.}
	\label{fig:framework}
\end{figure*}

In this paper, we propose JALAD, a joint accuracy and latency-aware deep network decoupling solution as shown in Fig. \ref{fig:framework}, to solve challenges including i) how to find the best partition of a deep structure; ii) how to deploy the component at an edge device that only has limited computation power; and finally iii) how to minimize the execution latency, consisting of the computation latency and transmission latency. Our contributions are summarized as follows.

$\rhd$ First, we study the structure and data flow in deep neural networks. Based on measurement studies on $4$ representative deep network models, including VGGNet \cite{VGG_Net} and ResNet \cite{Res_Net}, we reveal that there is significant room for compression for the data generated from intermediate layers. We find the intrinsic trade-off between in-layer \emph{data compression rate} and \emph{model accuracy}, and propose a quantization and image compression strategy that achieves a transmission data size of $1/2--1/5$ of the original image size, while maintaining the similar model accuracy. Furthermore, we introduce reinforcement learning based channel-wise feature removal to reduce the transmission data.

$\rhd$ Second, based on our insights above, we design a accuracy-and latency-aware decoupling solution. We build a model to capture the overall \emph{execution latency}, including edge computation latency, transmission latency and cloud computation latency. We formulate finding the ``right'' layer for decoupling as an ILP (Integer Linear Programming) problem to minimize the overall execution latency, subject to the required model accuracy. Our heuristic solution uses statistics from observed historical input data (e.g., images), and yields the layers for decoupling for different deep models in different bandwidth condition.

$\rhd$ Finally, we test our solution in real-world experiment and simulation, using the ILSVRC2012 \cite{ILSVRC15} dataset to verify the accuracy guarantee and latency speedup of our design. Compared with the conventional cloud-based solutions, our design can reduce the execution latency by up to $90\%$ in real-world experiment, while guaranteeing the accuracy loss within $10\%$.

The rest of the paper is organized as follows. We discuss related work in Sec.~\ref{sec:relatedwork}. We present our insights on deep network structure and the potential to decouple it in Sec.~\ref{sec:insights}, and present our detailed design and formulation in Sec.~\ref{sec:design}. We evaluate our strategy in Sec.~\ref{sec:evaluation} and conclude the paper in Sec.~\ref{sec:conclusion}.

\section{Framework and Challenge} \label{sec:insights}

In this section, we demonstrate our framework and the challenges in such deep network ``decoupling'' solution.

\subsection{Framework}

In Fig. \ref{fig:framework}, we present the framework of our joint accuracy- and latency-aware deep network decoupling design. We assume the deep neural networks share layered network structure. We decouple a deep neural network as follows: 1) We feed the raw data (e.g., an image) to its input layer and run some of its early layers on the edge device; 2) We compress the output data of the last layer at the edge device, and transfer it to the cloud, which runs the late layers of the neural network; 3) The partition layer will dynamically change over time according to the network bandwidth and the computation power of the edge device.

\subsection{Challenge: In-layer Data ``Amplification''}

\begin{figure}[htbp]
	\centering
	\vspace{-0.15cm}
	\includegraphics[width=\linewidth]{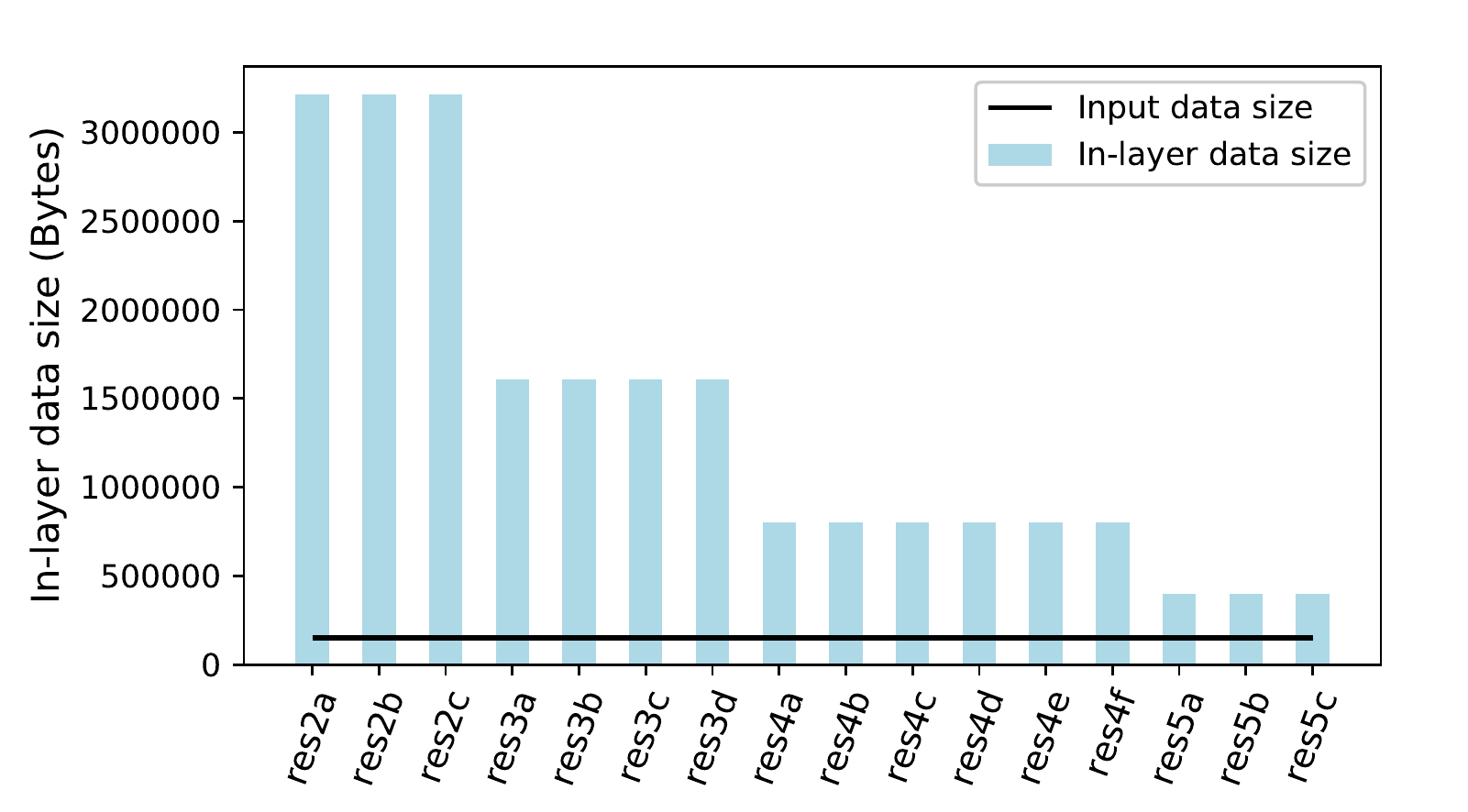}
	\caption{Data size at each decoupling point in ResNet, showing that the input data is amplified in in-layer feature maps.}
	\label{fig:in-layer}
	\vspace{-0.2cm}
\end{figure}

As suggested by \cite{Neurosurgeon}, simple decoupling of a deep neural network can hardly reduce its final execution latency in an edge-cloud deployment; because the \emph{in-layer} output data, to be transferred across the network, is usually much larger than the original input data, due to the existence of the \emph{convolutional} layers. For example, in our tests with pre-trained ResNet, the size of in-layer output data can be $20\times$ larger than the original input data in some early layers, as illustrated in Fig. \ref{fig:in-layer}. This also reveals why the partitioning generally ends up at very late layers in \cite{Neurosurgeon}.

\section{Detailed Design}
\label{sec:design}

In this section, we present the detailed design of our solution. We first define the decoupling points of a DNN,  then modeling all factors in the decoupling scheme. Notations used in the formulation is listed in Tab. ~\ref{tab:notations}

\begin{table}[htbp]
	\scriptsize
	\centering    
	\caption{Table of notations, $ i = 1,...,N $ represents layer's index of a DNN having $ N $ decoupling points}
	\label{tab:notations}
	\begin{tabular}{|l|l|}
		\hline
		Variable          & Description                                                                                                             \\ \hline
		$ T_{E_i} $         & Execution time of inference to layer $i$ on the edge device                                                             \\ \hline
		$ T_{C_i} $         & Execution time of inference from layer $i$ on the cloud server                                                          \\ \hline
		$ T_{\rm trans} $ & Time of transmission features from edge to the cloud                                                                    \\ \hline \hline
		$ N $            & Total count of layers in the current deep structure \\ \hline
		$ C $    & Total count of quantization bits \\ \hline
		$c_i$               & Feature values of layer $ i $ are quantized into $ c_i $ bits  \\ \hline
		$\mathcal{A}_i(c)$       & Accuracy drop when features of layer $i$ are quantized into $ c $ bits \\ \hline
		$ \mathcal{S}_i(c) $    & Data size of features of layer $ i $ when quantized into $ c $ bits \\ \hline
		$ \Delta \alpha $    & User defined accuracy drop boundary \\ \hline
		$ BW $ & Network bandwidth of the current application scenario \\ \hline
		$ x_{ic} $    & Decoupling decision on layer $ i $ and quantization bit $ c $ \\ \hline
	\end{tabular}
\end{table}

\subsection{Decoupling Points}

Not all nodes between layers can be used for decoupling. In traditional sequential DNNs like AlexNet \cite{AlexNet}, VGGNet, input signals flow straightforward layer by layer. But current deep models tents to introduce branchy network structures rather than sequential models (e.g., ResNet, GoogleNet \cite{GoogleNet} and InceptionNet \cite{InceptionNet}). To be general to all DNNs, we first define the decoupling points. In our solution, one res-unit in ResNet is regarded as one decoupling layer, equivalent to a convolution layer in sequential models. It means that when decoupling a sequential model, the granularity is layer-wise and when decoupling a branchy structure, the granularity is unit-wise.

\subsection{Accuracy-Aware Feature Compression}

Motivated by the fact that in-layer feature maps demonstrate strong sparsity, as the inter-layer feature maps illustrated in Fig. \ref{fig:framework}, {\color{black}{we try to compress the in-layer feature maps to reduce the data size while keeping the accuracy loss within our constraints and use several typical image recognition CNNs to test our framework.}}

\textbf{Conversion of float feature maps into small integers.} As the use of original float numbers in feature maps takes up a lot of the size, we propose to convert float feature maps into small integers, using a step conversion as below. Such quantization method has been proved feasible for compressing the parameters \cite{DeepCompression}, our experiments show that it also applies to compress feature values.

\begin{equation*}\label{equ: mapping}
y_i=\begin{cases}
\frac{(2^c-1) \big(x_i - \min(\mathbf{x})\big)}{\max(\mathbf{x})-\min(\mathbf{x})} & \text{if} \max(\mathbf{x})\ge 2^c \\
x_i & \text{otherwise},
\end{cases}
\end{equation*}

where $y_i$ is the converted integer, $x_i$ is the input float, $\mathbf{x}$ is the set of the original float numbers, and $c$ is the adaptive number of integer bits based on the network condition and the accuracy constraint. The rationale is to map the original float numbers in the feature map into the range of $[0,2^c)$ to reduce the size of in-layer feature maps. We use a step function because it approximates values without changing the value distribution in the feature maps.

\begin{figure*}[!th]
	\centering
	\includegraphics[width=\linewidth]{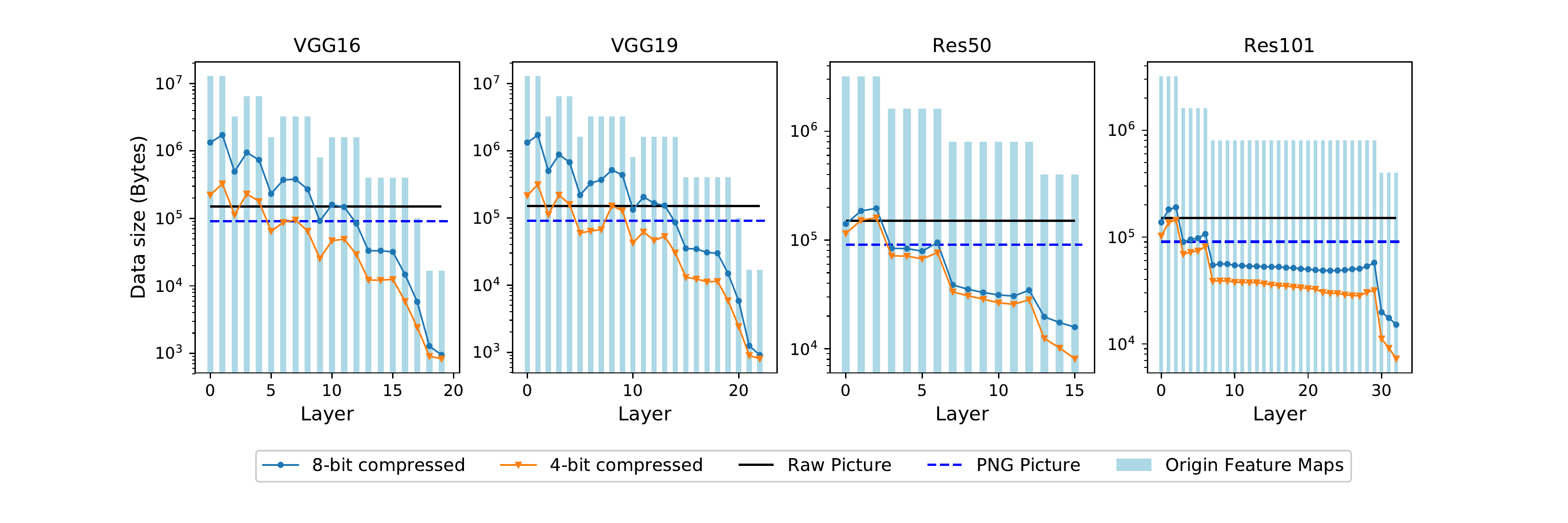}
	\vspace{-1cm}
	\caption{Compression performance for in-layer feature maps on different $ c $ setting.}
	\label{fig:compression_perf}
\end{figure*}

\textbf{Compression of integer feature maps.} Next, from our observation of the in-layer feature maps Fig. ~\ref{fig:framework}, the in-layer feature maps are highly sparse (many values are $ 0 $), therefore we can also benefit from conventional variable-length code. We introduce Huffman Coding to further compressed the quantized integer feature maps. To verify the compression performance, we compress the in-layer feature maps of 5000 samples from ILSVRC2012 validation set. As illustrated in Fig. \ref{fig:compression_perf}, we plot the in-layer data size for the original feature map and the compressed feature map, as well as the original input file size as references. We observe that our compression reduces the feature maps to $ 1/10 -- 1/100 $ of its original size, and enables us to partition at much more layers.

\subsection{Compressed accuracy and data size predictor}

\begin{figure}[!th]
	\centering
	\includegraphics[width=\linewidth]{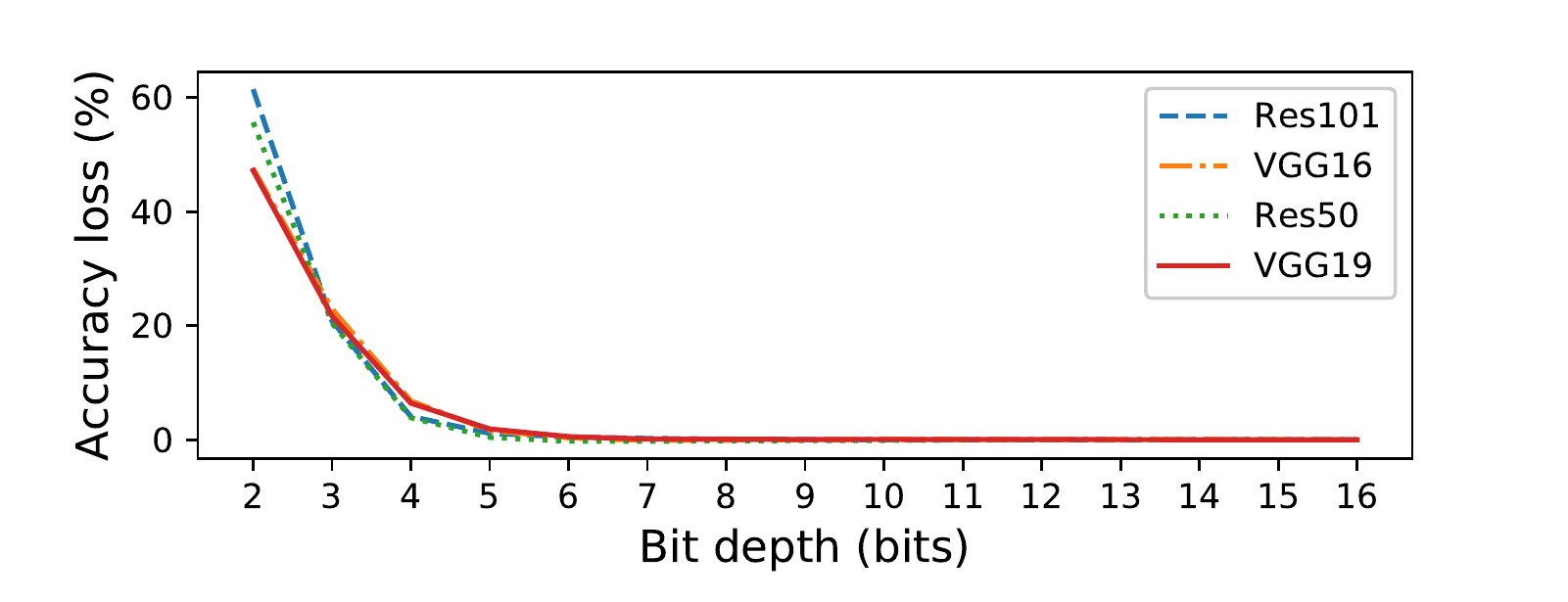}
	\caption{Accuracy loss $ \mathcal{A}(c) $ versus the number of bits used in compression $c$.}
	\label{fig:accuracy-vs-bits}
\end{figure}

\textbf{Trade-off between compression rate and accuracy.} Since the information loss in the feature map compression may affect the model's accuracy, we study the trade-off between the compression rate and model accuracy. As illustrated in Fig. \ref{fig:accuracy-vs-bits}, the curves represent the model accuracy loss threshold achieved on the same ILSVRC2012 dataset, versus the number of bits $c$ used in the compression. We observe that for these deep networks, $c\ge4$ already provides certain accuracy loss guarantee of $ 10\% $. 

\begin{figure}[!th]
	\centering
	\includegraphics[width=0.9\linewidth]{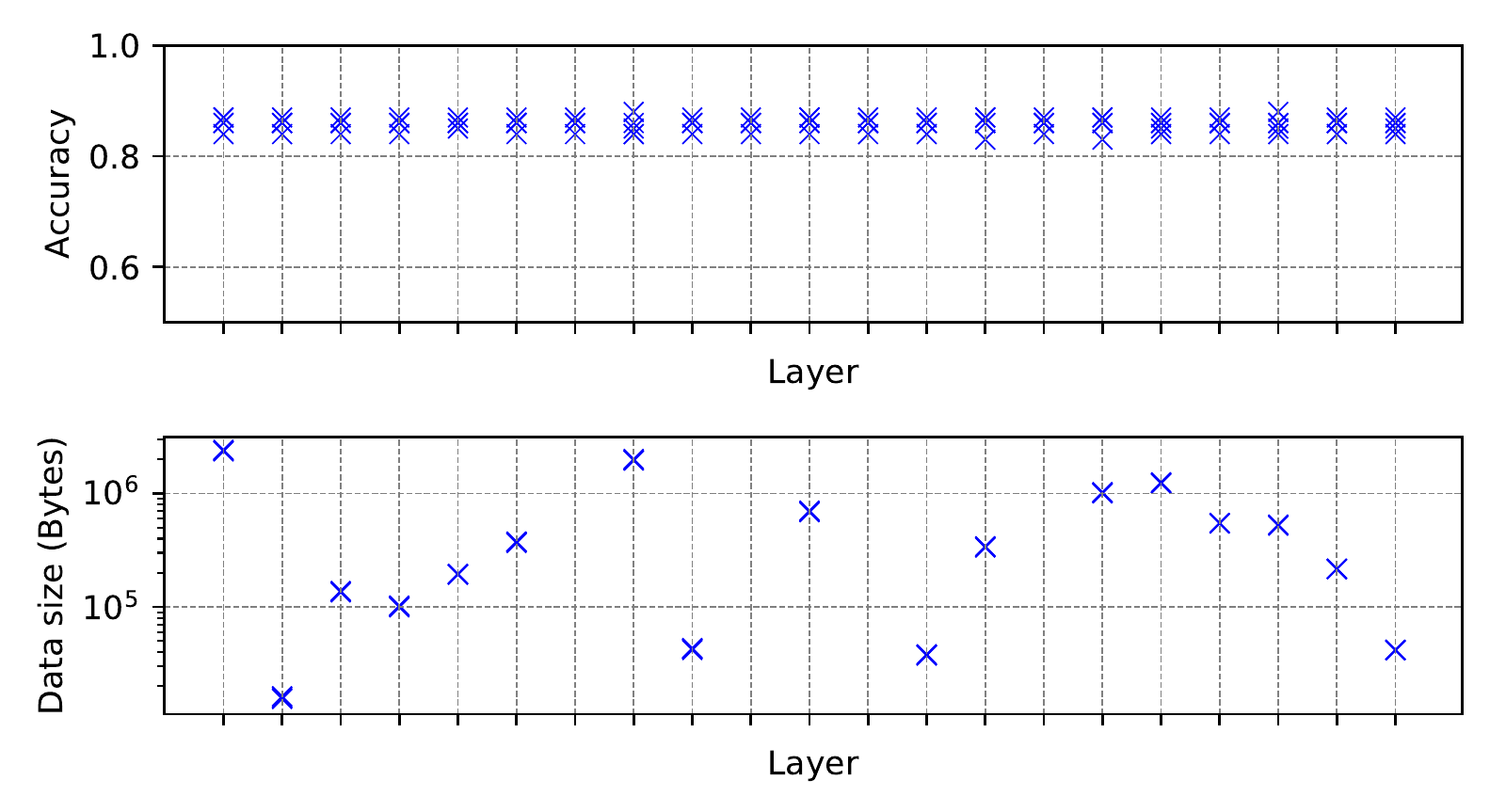}
	\caption{Accuracy and data size compressed results of different epochs are highly overlapped, therefore a specific compression setting will guarantee stable accuracy loss and data size in each layer. ($ c=8 $)}
	\label{fig:quantized_stability}
\end{figure}

\textbf{Train a predictor to predict accuracy loss and compressed size at a certain compression setting.} Feature quantization would result in accuracy loss, but the DNNs' prediction of an image is inexplicable yet, and the compressed data size is highly related to the input data. From Fig. ~\ref{fig:quantized_stability} we can observe that the accuracy loss and data size of a specific compression setting $ c $ is stable, therefore we can predict the current accuracy and compressed size based on historical statistics. We build a lookup table $ \mathcal{A}_i(c) $ to predict the accuracy loss and compressed data size $ \mathcal{S}_i(c) $ in a specific quantization bit $ c $.

The accuracy and compressed data size look-up tables are trained on ILSVRC2012, we iterate our compression scheme on all $ c \in C $. Since the compressed data size and accuracy presents stability in different test epochs, once the lookup table is built, we don't need a twice build up process. 

\begin{figure}[!th]
	\centering
	\includegraphics[width=\linewidth]{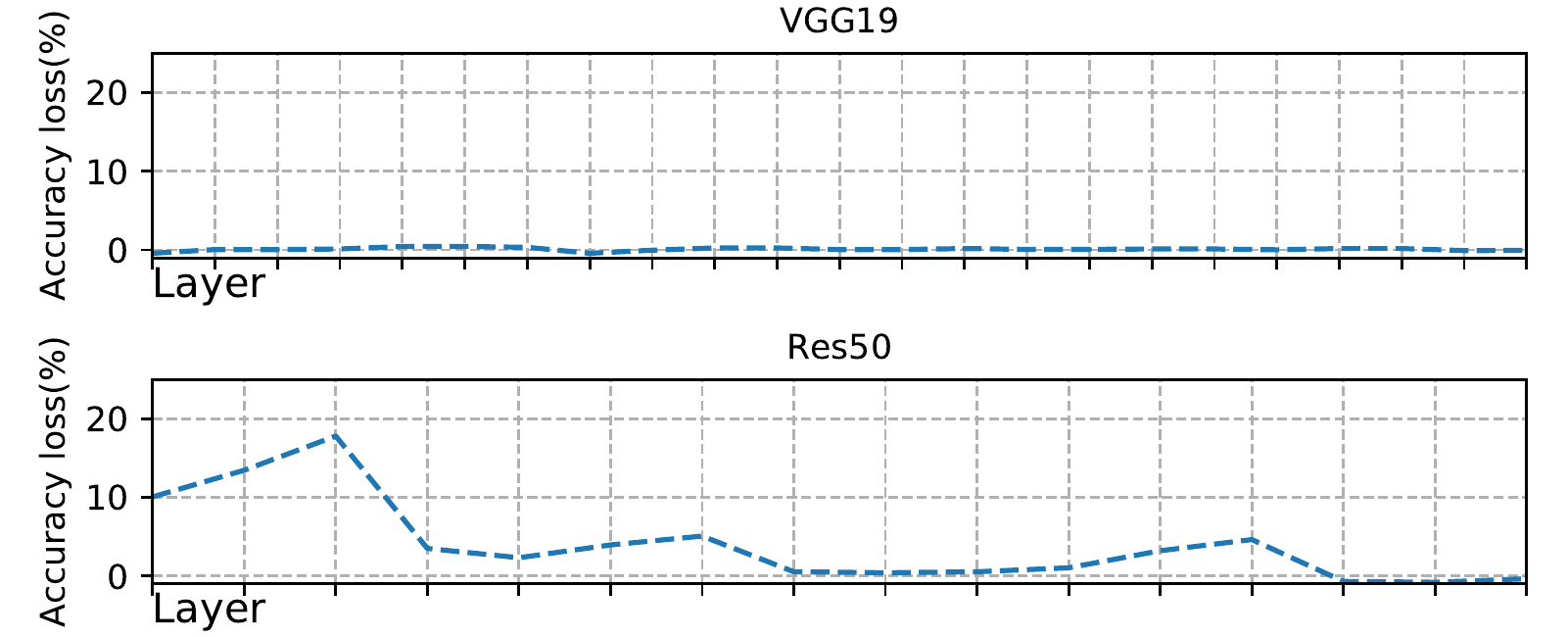}
	\caption{Accuracy loss $ \mathcal{A}_i(c), i=1,2,...,N $ for VGGNet and ResNet when compression takes place at different decoupling points (when $c=8$).}
	\label{fig:accuracy_reduce}
\end{figure}

One example of the accuracy predictor $ \mathcal{A}_i(c) $ is shown in Fig. ~\ref{fig:accuracy_reduce}, also, two examples of the compressed data size predictor when $ c=4, 8 $ are shown in Fig. ~\ref{fig:compression_perf}.

\subsection{Formulation of Decoupling and Execution Latency}

Based on the in-layer feature map compression, we then present the deep structure decoupling strategy, which aims to minimize the overall execution latency. 

Before finally get the decoupling decision and performance estimation, we have to get the latency for processing the input data. We use index $i \in {1,2,\ldots,N}$ to denote a layer, and $ N $ to denote the index of layers in a deep neural network. Without loss of generality, the input layer is indexed $1$. In a decoupling, we find an optimize partition $E$ (layers $1,2,\ldots,i^*$) and $ C $ (layers $i^*+1,i^*+2,\ldots,N$) at layer $i^*$ to minimize the overall execution latency, such that $L-E = C$, and layers in $E$ will execute at the edge device and layers in $C$ will execute at the cloud servers. The execution latency is compromised by the following components.

\textbf{Edge execution latency.} We denote $ T_{E_i} $ as the execution latency for running from layer $ 1 $ to layer $ i $. There are different approaches to estimate the execution time of a DNN. A parallel work \cite{Neurosurgeon} used a regression model to estimate the execution time of DNNs, \cite{paleo} and \cite{modeling_dnn_resource} tried to give analytic estimation of DNNs' execution time but failed to estimate fine-grained layer-level execution time.

Using one general model to estimate the execution time in all sorts of edge and cloud devices is difficult because of different commercial optimization policies. But for a specific device, the execution time tends to be stable on each execution, therefore we profile the execution time device-specifically, when our scheme is deployed, it would iteratively decouple the DNN on each layer and log the execution time of each layer to get $ \{T_{E_1}, T_{E_2},\ldots T_{E_N}\} $.

\textbf{Cloud execution latency.} The rest of layers are executed in the cloud, similarity, the cloud computation latency is denoted as $ T_{C_i} $ and we can get $ \{T_{C_1}, T_{C_2},\ldots T_{C_N}\} $ on the initialization stage of the deployment.

\textbf{Edge-cloud transmission latency.} In our decoupling design, another time cost is the edge-cloud latency caused by transferring the compressed feature maps to the cloud. The transmission latency $ T_{\rm trans} $ is related to the compressed data size and current network bandwidth. From the compressed data size predictor $ \mathcal{S}(c) $ we can look up the current data size to be transmitted. So, the edge-cloud latency if the partition takes place at layer $i$ and feature maps are quantized to $ c $ bits is calculated as: $ T_{\rm trans_{ic}} = \frac{\mathcal{S}_i(c)}{BW} $, where $ BW $ is the current network bandwidth.

\subsection{Decoupling Implementation}

From $ \mathcal{S}(c) $ we can look up the compressed feature data size, and predict the transmission latency $ T_{{\rm trans}_{ic}} $ on each layer $ i $ and quantized integer bit $ c $. With all $ T_{C_i}, T_{C_i} $ and $ T_{{\rm trans}_{ic}} $ calculated, on the decoupling decision phase, we are to fine the right decoupling layer index $ i^* $ and the right quantization bit $ c $ to minimize the overall joint-inference time $ Z $. Because layer index $ i $ and integer bit $ c $ are both discrete integer values, therefore we formulated the objective function as:

\begin{align*}
Z = \sum_{i=1}^N\sum_{c=1}^C T_{E_i} \cdot x_{ic} + \sum_{i=1}^N\sum_{c=1}^C T_{C_i} \cdot x_{ic} + \sum_{i=1}^N\sum_{c=1}^C T_{\rm trans} \cdot x_{ic} \\ 
\end{align*}

where $ x_{ic} $ is a binary variable indicating whether layer $ i $ is chosen as the decoupling layer, and the feature values of layer $ i $ are quantized into $ c $ bits.

In our solution, there should only be one decoupling point, therefore only one decoupling point $ i $ and only one quantization bit $ c $ should be chosen. Then we have a constraint to guarantee the uniqueness of the decision variable:

$$
\sum_{i=1}^N\sum_{c=1}^C x_{ic}=1 
$$

such that only one $ x_{ic} $ is set to $ 1 $ while all others are $ 0 $.

Also, in order to guarantee the minimum accuracy, we introduce a user-defined accuracy loss boundary $ \Delta\alpha $. From $ \mathcal{A}_i(c) $ we can have the statistical accuracy upon decoupling on layer $ i $ and features are quantized into $ c $ bits, therefore our optimization must guarantee the constraint:

$$
\sum_{i=1}^N\sum_{c=1}^C \mathcal{A}_i(c) \cdot x_{ic} \leq \Delta\alpha
$$

Finally, since $ \mathcal{A}_i(c) $ and $ \mathcal{S}_i(c) $ are pre-stored, and $ T_{E_i}, T_{C_i} $ are estimated in the initialization stage, as for the decoupling decision maker, $ T_{\rm trans}, T_{E_i}, T_{C_i} $ are just like ``constant" values. Combining the three parts of latency and the constraint above, we formulated the problem as an ILP problem:

\begin{align*}
\label{prob:ILP}
\min_{x_{ic}} \qquad &Z = \sum_{i=1}^N\sum_{c=1}^C T_{E_i} \cdot x_{ic} + \sum_{i=1}^N\sum_{c=1}^C T_{C_i} \cdot x_{ic} + \\
& \qquad \sum_{i=1}^N\sum_{c=1}^C T_{\rm trans} \cdot x_{ic} \\ 
{\rm s.t.} \qquad &\sum_{i=1}^N\sum_{c=1}^C x_{ic}=1 \\
& \sum_{i=1}^N\sum_{c=1}^C \mathcal{A}_i(c) \cdot x_{ic} \leq \Delta\alpha \\
{\rm vars} \qquad & x_{ic} \in \{0, 1\}
\end{align*}

In this way, we formulated finding the optimal decoupling layer $ i^* $ and quantization setting $ c $ as solving the above ILP problem. We further study the feasibility and time complexity to solve this problem.

\textbf{Feasibility.} From Fig. ~\ref{fig:accuracy_reduce} we can observe that quantizing feature values would not significantly affect model accuracy in some layers, especially in the last layer. Therefore when $ \Delta \alpha > 0 $ there would always be a feasible solution. The worst case is $ x_{NC} = 1 $, meaning no decoupling and quantization takes place at any of intermediate layers, and the above ILP problem becomes iteratively search in this case.

\textbf{Time complexity. }In the above ILP problem, the number of variable $ x_{ic} $ is equal to $ N\cdot C$ and therefore fixed and not too large, this fixed variable-amount ILP problem can be solved in polynomial time \cite{fixed_number_ilp}. In our test, solving an ILP problem of such scale in a desktop PC with Intel i7-6800K CPU takes $ 1.77 $ ms.

Our algorithm estimates $ T_{\rm trans} $ by assigning the current network bandwidth $ BW $ and refer to $ \mathcal{S}(c) $, solving the ILP problem to find the partition $ i^* $ and the appropriate number of integer bits $ c $ with the minimum latency that satisfies the accuracy constraint $ \Delta\alpha $. Then the edge and cloud will ``synchronize'' using the new decoupling. Generally, our design re-decouples the deep neural network upon the edge-cloud network change, thus in different edge and network conditions, our solution can adaptively use different decoupling schemes.

\section{Evaluation}
\label{sec:evaluation}

In this section, we use both real-world experiments and controlled simulations to evaluate the effectiveness of our design.

\subsection{Experiment Setup}

\emph{Real-world Experiment.} We carry out real-world experiments based on a cloud server with $1$ NVIDIA 1080ti GPU, and an edge device with $1$ NVIDIA Quadro K620 GPU. We control the network bandwidth between them and measure the accuracy and latency performance, and choosing different accuracy loss boundary $ \Delta\alpha $ to observe JALAD's decoupling decision. We use ILSVRC2012 validation set as the testing input image data.

\emph{Simulation Experiment.} Since we don't have various kinds of devices for the evaluation of computation impact, to explore the impact of the computation power of different edge devices, we further run controlled simulations.

We denoted that $ Q({\bf x}) $ as the amount of FMACs (floating-point multiply-add calculations) of layers set $ {\bf x} $, therefore all FMACs run on edge and cloud is $ Q(E) $ and $ Q(C) $, and denoted that a device's FLOPS (floating-point operations per second) as $ F $. By simplifying $T_E$ and $T_C$ as linear functions $ w_e\cdot \frac{Q(E)}{F_E} $ and $ w_c\cdot \frac{Q(C)}{F_C} $, where $w_e$ and $w_c$ are fitting parameters to regress the real execution time. Since FLOPS is an important performance metric of a device, and FMACs have been proven to take up more than $ 90\% $ of the execution time \cite{modeling_dnn_resource}, this approximation is still credible. In our experiments, we let $F_C=12$TFLOPs; and $F_E=2$TFLOPs when the edge is a high-performance device according to NVIDIA Tegra X2 and $F_E = 300$GFLOPs when it is a low-performance device, based on NVIDIA Tegra K1. The average edge-cloud bandwidth $V$ is set to $ 1 $MBps. $w_e$ and $w_c$ is set to $ 1.1176 $ and $ 2.1761 $ respectively in our simulation, which is exported from experimental regression on NVIDIA 1080ti with $ F=10.5$ TFLOPS.

\emph{Baselines.} We use the following baselines in our evaluation: 1) Origin2Cloud: Which transfers the original raw images (8-bit RGB pictures) to the cloud server that runs the deep neural networks; and 2) JPEG2Cloud, which transfers lossy compressed images in JPEG files to the cloud server and finish the entire inference process on the cloud, which is the conventional cloud-based AI approach currently.

\emph{Testing Models and Metrics.} We decoupled VGGNet (16 layers and 19 layers version), and ResNet (50 layers and 101 layers version) to evaluate JALAD's execution speedup and accuracy performance. In particular, we use $100$ samples from ILSVRC2012 in each iteration and calculate the average latency in $20$ iterations.

\subsection{Execution Latency Speedup}

First, we study the execution speedup of JALAD against conventional pure cloud strategies, which is defined as the fraction of baselines' execution latency over the JALAD execution latency.

As shown in Table \ref{tab:real_latency}, for the $300$ KBps edge-cloud bandwidth scenario, we observe that the latency speedup for VGGNet against PNG2Cloud (resp. Origin2Cloud) can be up to $ 3.6\times $ (resp. $6.0\times$); and we achieve $ 5.6\times $ (resp. $ 9.3\times $) latency speedup against PNG2Cloud (resp. Origin2Cloud) on average in ResNet. When the bandwidth is increased to $1$ MBps, JALAD lets more layers execute on the cloud, taking the advantage of smaller transmission latency, with an average $1.6\times$ (resp. $2.5\times$) execution speedup against PNG2Cloud (resp. Origin2Cloud) across 4 models. In this test, the accuracy loss threshold $ \Delta\alpha $ is set to $ 10\% $


\begin{table}[!hb]
	\centering
	\small
	\caption{Execution speedup on different network conditions.}
	\label{tab:real_latency}
	\begin{tabular}{@{}lll@{}}
		\toprule
		& 1MBps & 300KBps \\ \midrule
		& PNG2Cloud/Origin2Cloud & PNG2Cloud/Origin2Cloud \\ \midrule
		VGG16 & 1.4$\times$/2.2$\times$ & 3.6$\times$/6.0$\times$ \\
		VGG19 & 1.1$\times$/1.7$\times$ & 3.0$\times$/4.9$\times$ \\
		Res50 & 2.3$\times$/3.7$\times$ & 7.2$\times$/11.7$\times$ \\
		Res101 & 1.5$\times$/2.3$\times$ & 4.3$\times$/6.9$\times$ \\ \bottomrule
	\end{tabular}
\end{table}



\subsection{Impact of Accuracy Threshold}

\begin{figure}[t]
	\begin{minipage}[b]{1\linewidth}
		\centering
		\centerline{\epsfig{figure=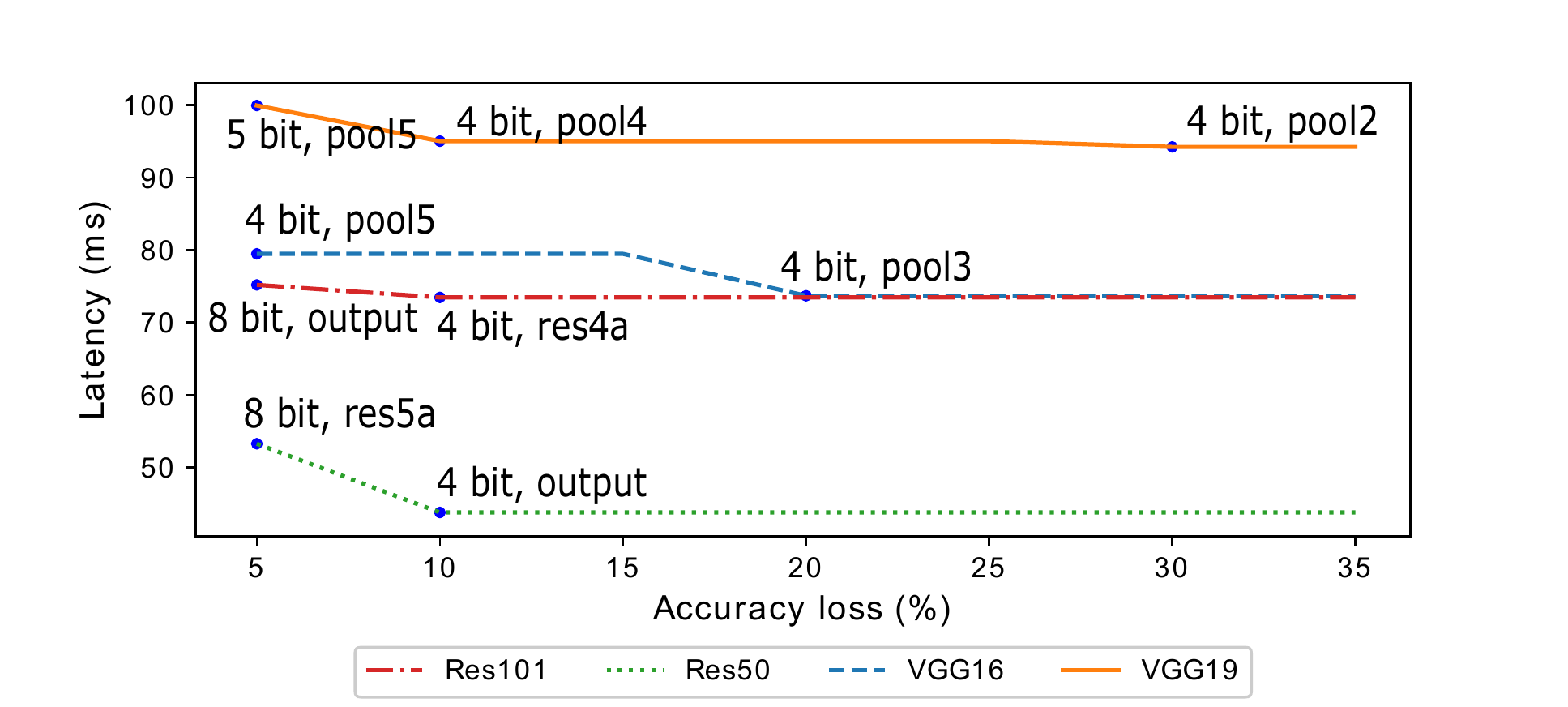, width=9cm}}
	\end{minipage}
	\caption{Accuracy versus latency}
	\label{fig:acc_latency_tradeoff}
\end{figure}

Next, we study the impact of our design on the model accuracy. We choose different accuracy threshold $ \Delta\alpha $ to test JALAD's latency performance and plot the average execution latency and decoupling decision in Figure \ref{fig:acc_latency_tradeoff}. We observe that as the threshold increases, our design can achieve better latency gain. The reason is that JALAD can either change the decoupling layer or cast the in-layer feature maps into lower bit-depth (which means transmitting fewer bytes) to achieve lower latency.

\subsection{Impact of Cloud-Edge Bandwidth Variation}

\begin{figure}[t]
	\begin{minipage}[b]{1\linewidth}
		\centering
		\centerline{\epsfig{figure=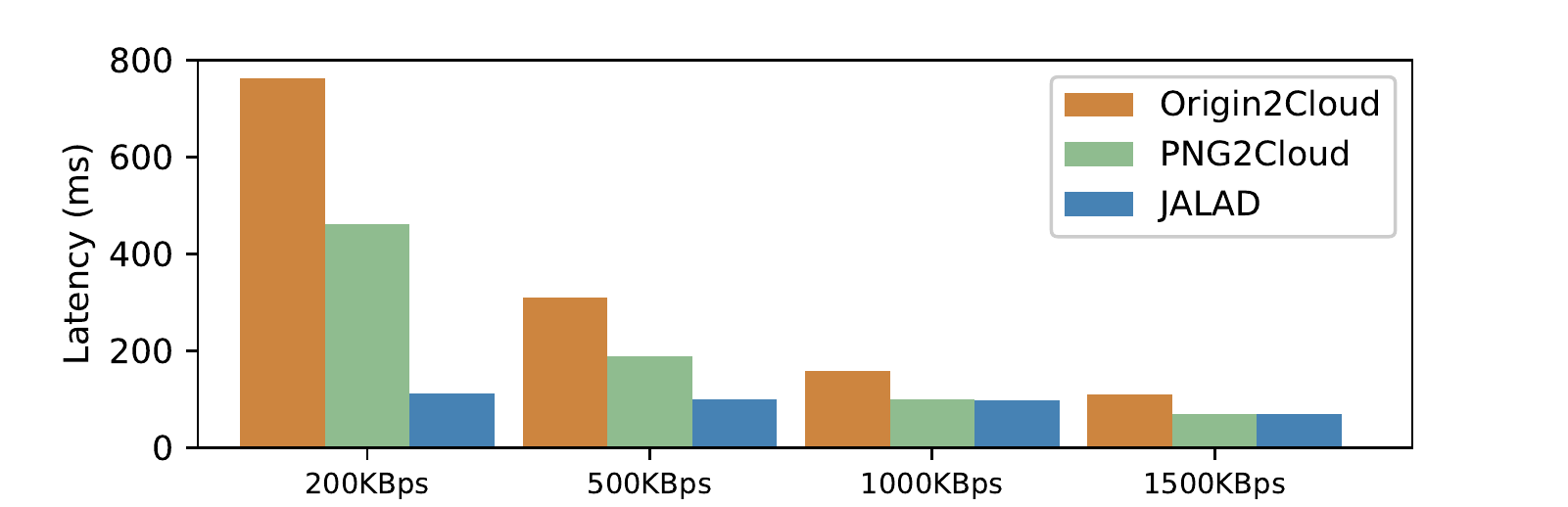, width=9cm}}
	\end{minipage}
	\caption{Execution speedup under different edge-cloud network conditions.}
	\label{fig:jalad_stability}
\end{figure}

Figure \ref{fig:jalad_stability} shows the impact of changing edge-cloud bandwidth, by plotting the execution latencies achieved by different decoupling strategies under different edge-cloud bandwidths. We observe that JALAD remains a stable low latency by adaptively changing the decoupling strategy, while the baseline strategies are significantly affected by the upload capacity. Specifically, when the network condition is good (i.e. 1.5MBps in our experiment), JALAD tends to upload the raw PNG images to the cloud, thus we have the same latency performance with the PNG2Cloud solution, and in some applications(e.g., auto driving vehicle) that the network condition is poor, our solution can achieve higher speed-up gain.

\subsection{Impact of Edge Device Computation Power}

In order to further evaluate the impact of edge device computation power, we carry out simulation experiments. As illustrated in Table \ref{tab:diff_edge}, in the simulation scenario with $ 1 $ MBps bandwidth, we observe that JALAD achieves more execution speedup gain under the high-performance edge device NVIDIA Tegra X2, and the average speedup is $7.6\times$ (resp. $12.6\times$) against the PNG2Cloud (resp. Origin2Cloud) for the four representative deep networks; We also notice that the execution latency cannot benefit from decouple when the computation power of an edge device is limited, for certain networks including VGGNet. The reason is that the limited edge computation power hinders JALAD to explore possible decoupling layers.

\begin{table}[!hb]
	\centering
	\small
	\caption{Execution speedup on different edge performance.}
	\begin{tabular}{@{}lll@{}}
		\toprule
		& NVIDIA Tegra K1 & NVIDIA Tegra X2     \\ \midrule
		& PNG2Cloud/Origin2Cloud & PNG2Cloud/Origin2Cloud \\ \midrule
		VGG16  & 1.0$\times$ / 1.5$\times$     & 3.4$\times$ / 5.5$\times$ \\
		VGG19  & 1.0$\times$ / 1.5$\times$       & 2.9$\times$ / 4.7$\times$   \\
		Res50  & 2.2$\times$ / 3.7$\times$     & 15.1$\times$ / 25.1$\times$  \\
		Res101 & 1.4$\times$ / 2.3$\times$       & 9.0$\times$ / 14.9$\times$  \\ \bottomrule
	\end{tabular}
	\label{tab:diff_edge}
\end{table}

\section{Related Work}
\label{sec:relatedwork}


%
%
%
%
%
%

Deep neural network has become the \emph{de facto} structure for today's practical machine learning model selection \cite{taigman2014deepface}, thanks to its simplicity and effectiveness. In order to deploy the tons of pre-trained deep neural networks and run them on different devices, researchers propose the following deep network model deployment solutions. 

Attracted by the elasticity in computing power and flexible collaboration, hierarchically distributed computing structures(e.g., cloud computing, fog computing, edge computing) naturally becomes the choice for supporting deep-structure-based services and applications \cite{Fog_computing_survey,edge_computing_survey,bonomi2012fog,skala2015scalable,stoica2017berkeley}. Considering the deployment location for model, the-state-of-art approaches could be divided into the three classes.

\emph{Cloud-based deployment}: Conventionally, most of  today's deep neural network usually deployed on dedicated servers in the datacenter \cite{skala2015scalable}. Users usually have to upload a large amount of original data (e.g., images) to the servers, causing the latency. To reduce the latency, \cite{chen2015glimpse} proposed a bandwidth efficiency object tracking system by dropping video frames from the raw video to the cloud; Jain et al.~\cite{jain2015overlay} proposed to make use of the blurred frames so as to reduce upload load. The limitation of conventional cloud-based studies is that to some extent they have to upload the orignal data, which causes large latency.

\emph{Edge-based deployment}: An alternative is to deploy deep neural networks at the edge of network \cite{ha2013just,ha2014towards}. In \cite{lane2015deepear,lane2016deepx}, Lane et al.~first explored the feasibility of running deep neural networks on mobile devices for audio-sensing applications. Hu et al.~\cite{Edge_LBP} proposed face identification deployment framework based on a fog computing infrustructure. Later, Huynh et al.~\cite{huynh2017deepmon} optimized the execution of deep learning models on mobile GPUs. Though close to end users, the pure edge-based deployment is limited by its computational power, making it hard to run current general neural network models, and running a compressed model lacks of elasticity on customizing for different accuracy demand.

Despite the edge-based approaches could reducing the end to end latency, the edge(mobile device) still falls short in addressing the high energy consumption and computation workload challenges.

\emph{Hybrid deployment}: Alternatively, some studies focus on offloading the tasks to the edge server instead of running the DNNs on the mobile devices, \cite{satyanarayanan2009case} explpited virtual machine technique to offload the customized service on a one wireless-hop away server called ``cloudlet" to speed up service. \cite{ha2013just,ha2014towards} used ``cloudlet" for speech and face recognition application. To this end, hybrid deployment was proposed to using both central cloud servers and edge devices for the execution. Traditional solution to deal with the trade-off is offloading computation from client\cite{MAUI, mobile_cloud_offloading, collaborative_mobile_offloading}, but these approaches are designed for generalized application, did not focus on DNN's special structure. 

For current applications with deep neural network structure, some partition scheme\cite{Edge_LBP, Offloading_image_classification} tried to extract features locally to assist the classification DNN in the cloud. But these schemes did not talk about partition on DNN structure. \cite{Han2016} implemented a streaming analysis system that balances the classification accuracy and resource demand by scheduling model variants between resource-constrained devices and cloud. \cite{Neurosurgeon} proposed to partition the deep network layers so that they run at the mobile devices and cloud, respectively. \cite{DDNN} further proposed to partition the deep network onto the cloud, fog computing devices (e.g., a basestation) and users' mobile devices. \cite{ILP_Decoupling} presented a solution to decouple a deep structure into 3 segments without in-layer feature compression.

The limitation of previous studies is that they fail to take the compression of in-layer data into consideration, causing the latency gain very limited even with a joint edge and cloud deployment.

However, these former proposal took into account only latency measurement and raw data quantity between layers, not taking the sparsity in the feature maps into account, thus their partition point frequently falls on the first or the last layer of a DNN structure in their experiments, which makes their divided model turns to be cloud-only or client-only scheme and thus less practical. Another work \cite{DeepDecision} proposed to adaptively upload compressed data or deploy compressed model locally to jointly execute, but they didn't look into and modify the deep model itself.

\section{Concluding Remarks} \label{sec:conclusion}

As many machine learning services and applications are based on deep neural networks, developing their execution infrastructure is of great importance. In this paper, we study the trade-off between the model accuracy and execution latency when a deep neural network is decoupled to execute on an edge device and the cloud, respectively. Our study reveals that such decoupling has potential to reduce the overall execution latency, we propose an accuracy-aware strategy for in-layer feature map compression to enable the decoupling. We further formulate the deep structure decoupling as an optimization problem to minimize the overall execution latency, with a guaranteed accuracy constraint. Our real-world experiments based on 4 representative deep neural networks demonstrate that our design can speed up the execution while guaranteeing the accuracy loss within a user-defined boundary.


\begin{thebibliography}{10}
\providecommand{\url}[1]{#1}
\csname url@samestyle\endcsname
\providecommand{\newblock}{\relax}
\providecommand{\bibinfo}[2]{#2}
\providecommand{\BIBentrySTDinterwordspacing}{\spaceskip=0pt\relax}
\providecommand{\BIBentryALTinterwordstretchfactor}{4}
\providecommand{\BIBentryALTinterwordspacing}{\spaceskip=\fontdimen2\font plus
\BIBentryALTinterwordstretchfactor\fontdimen3\font minus
  \fontdimen4\font\relax}
\providecommand{\BIBforeignlanguage}[2]{{%
\expandafter\ifx\csname l@#1\endcsname\relax
\typeout{** WARNING: IEEEtran.bst: No hyphenation pattern has been}%
\typeout{** loaded for the language `#1'. Using the pattern for}%
\typeout{** the default language instead.}%
\else
\language=\csname l@#1\endcsname
\fi
#2}}
\providecommand{\BIBdecl}{\relax}
\BIBdecl

\bibitem{google-tpu}
``Google supercharges machine learning tasks with tpu custom chip,
  https://cloudplatform.googleblog.com/2016/05/google-supercharges-machine-learning-tasks-with-custom-chip.html,
  accessed: 2017-01.''

\bibitem{apple-siri}
``Apple moves to third-generation siri back-end, built on open-source mesos
  platform, http://9to5mac.com/2015/04/27/siri-backend-mesos/, accessed:
  2016-08.''

\bibitem{index520862global}
C.~V.~N. Index, ``Global mobile data traffic forecast update 2014--2019. white
  paper c11-520862,'' \emph{Available on http://www. cisco.
  com/c/en/us/solutions/collateral/service-provider/visual-networking-index-vni/white\_paper\_c11-520862.
  html}.

\bibitem{roman2016mobile}
R.~Roman, J.~Lopez, and M.~Mambo, ``Mobile edge computing: a survey and
  analysis of security threats and challenges,'' \emph{arXiv preprint
  arXiv:1602.00484}, 2016.

\bibitem{structured_pruning}
S.~Anwar, K.~Hwang, and W.~Sung, ``Structured pruning of deep convolutional
  neural networks,'' \emph{ACM Journal on Emerging Technologies in Computing
  Systems (JETC)}, vol.~13, no.~3, p.~32, 2017.

\bibitem{pruning_han}
S.~Han, J.~Pool, J.~Tran, and W.~Dally, ``Learning both weights and connections
  for efficient neural network,'' in \emph{Advances in neural information
  processing systems}, 2015, pp. 1135--1143.

\bibitem{prun_quanti}
S.~Han, H.~Mao, and W.~J. Dally, ``A deep neural network compression pipeline:
  Pruning, quantization, huffman encoding,'' \emph{arXiv preprint
  arXiv:1510.00149}, vol.~10, 2015.

\bibitem{quantization}
Y.~Gong, L.~Liu, M.~Yang, and L.~Bourdev, ``Compressing deep convolutional
  networks using vector quantization,'' \emph{arXiv preprint arXiv:1412.6115},
  2014.

\bibitem{quantize_3bit}
K.~Hwang and W.~Sung, ``Fixed-point feedforward deep neural network design
  using weights+ 1, 0, and- 1,'' in \emph{Signal Processing Systems (SiPS),
  2014 IEEE Workshop on}.\hskip 1em plus 0.5em minus 0.4em\relax IEEE, 2014,
  pp. 1--6.

\bibitem{quantize}
S.~Anwar, K.~Hwang, and W.~Sung, ``Fixed point optimization of deep
  convolutional neural networks for object recognition,'' in \emph{Acoustics,
  Speech and Signal Processing (ICASSP), 2015 IEEE International Conference
  on}.\hskip 1em plus 0.5em minus 0.4em\relax IEEE, 2015, pp. 1131--1135.

\bibitem{Neurosurgeon}
Y.~Kang, J.~Hauswald, C.~Gao, A.~Rovinski, T.~Mudge, J.Mars, and L.~Tang,
  ``Neurosurgeon: collaborative intelligence between the cloud and mobile
  edge,'' in \emph{ASPLOS}.\hskip 1em plus 0.5em minus 0.4em\relax ACM, 2017,
  pp. 615--629.

\bibitem{VGG_Net}
K.~Simonyan and A.~Zisserman, ``Very deep convolutional networks for
  large-scale image recognition,'' \emph{arXiv preprint arXiv:1409.1556}, 2014.

\bibitem{Res_Net}
K.~He, X.~Zhang, S.~Ren, and J.~Sun, ``Deep residual learning for image
  recognition,'' in \emph{CVPR}, 2016, pp. 770--778.

\bibitem{ILSVRC15}
O.~Russakovsky, J.~Deng, H.~Su, J.~Krause, S.~Satheesh, S.~Ma, Z.~Huang,
  A.~Karpathy, A.~Khosla, M.~Bernstein \emph{et~al.}, ``Imagenet large scale
  visual recognition challenge,'' \emph{International Journal of Computer
  Vision}, vol. 115, no.~3, pp. 211--252, 2015.

\bibitem{AlexNet}
A.~Krizhevsky, I.~Sutskever, and G.~E. Hinton, ``Imagenet classification with
  deep convolutional neural networks,'' in \emph{NIPS}, 2012, pp. 1097--1105.

\bibitem{GoogleNet}
C.~Szegedy, W.~Liu, Y.~Jia, P.~Sermanet, S.~Reed, D.~Anguelov, D.~Erhan,
  V.~Vanhoucke, and A.~Rabinovich, ``Going deeper with convolutions,'' in
  \emph{Proceedings of the IEEE conference on computer vision and pattern
  recognition}, 2015, pp. 1--9.

\bibitem{InceptionNet}
C.~Szegedy, V.~Vanhoucke, S.~Ioffe, J.~Shlens, and Z.~Wojna, ``Rethinking the
  inception architecture for computer vision,'' in \emph{Proceedings of the
  IEEE conference on computer vision and pattern recognition}, 2016, pp.
  2818--2826.

\bibitem{DeepCompression}
S.~Han, H.~Mao, and W.~J. Dally, ``Deep compression: Compressing deep neural
  networks with pruning, trained quantization and huffman coding,'' \emph{arXiv
  preprint arXiv:1510.00149}, 2015.

\bibitem{paleo}
H.~Qi, E.~R. Sparks, and A.~Talwalkar, ``Paleo: A performance model for deep
  neural networks,'' 2016.

\bibitem{modeling_dnn_resource}
Z.~Lu, S.~Rallapalli, K.~Chan, and T.~La~Porta, ``Modeling the resource
  requirements of convolutional neural networks on mobile devices,'' in
  \emph{Proceedings of the 2017 ACM on Multimedia Conference}.\hskip 1em plus
  0.5em minus 0.4em\relax ACM, 2017, pp. 1663--1671.

\bibitem{fixed_number_ilp}
H.~W. Lenstra~Jr, ``Integer programming with a fixed number of variables,''
  \emph{Mathematics of operations research}, vol.~8, no.~4, pp. 538--548, 1983.

\bibitem{taigman2014deepface}
Y.~Taigman, M.~Yang, M.~Ranzato, and L.~Wolf, ``Deepface: closing the gap to
  human-level performance in face verification,'' in \emph{CVPR}, 2014, pp.
  1701--1708.

\bibitem{Fog_computing_survey}
S.~Yi, C.~Li, and Q.~Li, ``A survey of fog computing: concepts, applications
  and issues,'' pp. 37--42, 2015.

\bibitem{edge_computing_survey}
W.~Shi, J.~Cao, Q.~Zhang, Y.~Li, and L.~Xu, ``Edge computing: vision and
  challenges,'' \emph{IEEE Internet of Things Journal}, vol.~3, no.~5, pp.
  637--646, 2016.

\bibitem{bonomi2012fog}
F.~Bonomi, R.~Milito, J.~Zhu, and S.~Addepalli, ``Fog computing and its role in
  the internet of things,'' in \emph{Proceedings of the first edition of the
  MCC workshop on Mobile cloud computing}.\hskip 1em plus 0.5em minus
  0.4em\relax ACM, 2012, pp. 13--16.

\bibitem{skala2015scalable}
K.~Skala, D.~Davidovic, E.~Afgan, I.~Sovic, and Z.~Sojat, ``Scalable
  distributed computing hierarchy: cloud, fog and dew computing,'' \emph{Open
  Journal of Cloud Computing (OJCC)}, vol.~2, no.~1, pp. 16--24, 2015.

\bibitem{stoica2017berkeley}
I.~Stoica, D.~Song, R.~A. Popa, D.~A. Patterson, M.~W. Mahoney, R.~H. Katz,
  A.~D. Joseph, M.~Jordan, J.~M. Hellerstein, J.~Gonzalez \emph{et~al.}, ``A
  berkeley view of systems challenges for ai,'' 2017.

\bibitem{chen2015glimpse}
T.~Y.-H. Chen, L.~Ravindranath, S.~Deng, P.~Bahl, and H.~Balakrishnan,
  ``Glimpse: continuous, real-time object recognition on mobile devices,'' in
  \emph{13th ACM Conference on Embedded Networked Sensor Systems}.\hskip 1em
  plus 0.5em minus 0.4em\relax ACM, 2015, pp. 155--168.

\bibitem{jain2015overlay}
P.~Jain, J.~Manweiler, and R.~Roy~Choudhury, ``Overlay: practical mobile
  augmented reality,'' in \emph{MobiSys}.\hskip 1em plus 0.5em minus
  0.4em\relax ACM, 2015, pp. 331--344.

\bibitem{ha2013just}
K.~Ha, P.~Pillai, W.~Richter, Y.~Abe, and M.~Satyanarayanan, ``Just-in-time
  provisioning for cyber foraging,'' in \emph{MobiSys}.\hskip 1em plus 0.5em
  minus 0.4em\relax ACM, 2013, pp. 153--166.

\bibitem{ha2014towards}
K.~Ha, Z.~Chen, W.~Hu, W.~Richter, P.~Pillai, and M.~Satyanarayanan, ``Towards
  wearable cognitive assistance,'' in \emph{MobiSys}.\hskip 1em plus 0.5em
  minus 0.4em\relax ACM, 2014, pp. 68--81.

\bibitem{lane2015deepear}
N.~D. Lane, P.~Georgiev, and L.~Qendro, ``Deepear: robust smartphone audio
  sensing in unconstrained acoustic environments using deep learning,'' in
  \emph{2015 ACM International Joint Conference on Pervasive and Ubiquitous
  Computing}.\hskip 1em plus 0.5em minus 0.4em\relax ACM, 2015, pp. 283--294.

\bibitem{lane2016deepx}
N.~D. Lane, S.~Bhattacharya, P.~Georgiev, C.~Forlivesi, L.~Jiao, L.~Qendro, and
  F.~Kawsar, ``Deepx: a software accelerator for low-power deep learning
  inference on mobile devices,'' in \emph{15th ACM/IEEE International
  Conference on Information Processing in Sensor Networks}.\hskip 1em plus
  0.5em minus 0.4em\relax IEEE, 2016, pp. 1--12.

\bibitem{Edge_LBP}
P.~Hu, H.~Ning, T.~Qiu, Y.~Zhang, and X.~Luo, ``Fog computing-based face
  identification and resolution scheme in internet of things,'' \emph{IEEE
  Transactions on Industrial Informatics}, vol.~13, no.~4, pp. 1910 -- 1920,
  2017.

\bibitem{huynh2017deepmon}
L.~N. Huynh, Y.~Lee, and R.~K. Balan, ``Deepmon: mobile gpu-based deep learning
  framework for continuous vision applications,'' in \emph{MobiSys}.\hskip 1em
  plus 0.5em minus 0.4em\relax ACM, 2017, pp. 82--95.

\bibitem{satyanarayanan2009case}
M.~Satyanarayanan, P.~Bahl, R.~Caceres, and N.~Davies, ``The case for vm-based
  cloudlets in mobile computing,'' \emph{IEEE pervasive Computing}, vol.~8,
  no.~4, 2009.

\bibitem{MAUI}
E.~Cuervo, A.~Balasubramanian, and D.~ki~Cho~et al., ``Maui: making smartphones
  last longer with code offload,'' in \emph{MobiSys}.\hskip 1em plus 0.5em
  minus 0.4em\relax ACM, 2010.

\bibitem{mobile_cloud_offloading}
W.~Zhang, Y.~Wen, and H.-H. Chen, ``Toward transcoding as a service:
  energy-efficient offloading policy for green mobile cloud,'' \emph{IEEE
  Network}, vol.~28, no.~6, pp. 67--73, 2014.

\bibitem{collaborative_mobile_offloading}
W.~Zhang, Y.~Wen, and D.~O. Wu, ``Collaborative task execution in mobile cloud
  computing under a stochastic wireless channel,'' \emph{IEEE Transactions on
  Wireless Communications}, vol.~14, no.~1, pp. 81--93, 2015.

\bibitem{Offloading_image_classification}
J.~Hauswald, T.~Manville, Q.~Zheng, R.~Dreslinski, C.~Chakrabarti, and
  T.~Mudge, ``A hybrid approach to offloading mobile image classification,'' in
  \emph{ICASSP}, 2014, pp. 8375--8379.

\bibitem{Han2016}
S.~Han, H.~Shen, M.~Philipose, S.~Agarwal, A.~Wolman, and A.~Krishnamurthy,
  ``Mcdnn: an approximation-based execution framework for deep stream
  processing under resource constraints,'' in \emph{MobiSys}.\hskip 1em plus
  0.5em minus 0.4em\relax ACM, 2016, pp. 123--136.

\bibitem{DDNN}
S.~Teerapittayanon, B.~McDanel, and H.~Kung, ``Distributed deep neural networks
  over the cloud, the edge and end devices,'' in \emph{ICDCS}.\hskip 1em plus
  0.5em minus 0.4em\relax IEEE, 2017, pp. 328--339.

\bibitem{ILP_Decoupling}
A.~E. Eshratifar and M.~Pedram, ``Energy and performance efficient computation
  offloading for deep neural networks in a mobile cloud computing
  environment,'' in \emph{Proceedings of the 2018 on Great Lakes Symposium on
  VLSI}.\hskip 1em plus 0.5em minus 0.4em\relax ACM, 2018, pp. 111--116.

\bibitem{DeepDecision}
X.~Ran, H.~Chen, X.~Zhu, Z.~Liu, and J.~Chen, ``Deepdecision: A mobile deep
  learning framework for edge video analytics,'' \emph{INFOCOM}, 2018.

\end{thebibliography}

%

\end{document}